\documentclass{aa}
\usepackage{graphics,latexsym,amssymb,times,psfig,amsmath}
\def\gsim{ \lower .75ex \hbox{$\sim$} \llap{\raise .27ex \hbox{$>$}} } 
\def\lsim{ \lower .75ex\hbox{$\sim$} \llap{\raise .27ex \hbox{$<$}} } 
\def\ghi{$E_{\rm peak}-E_{\gamma}$}
\def\lz{ $E_{\rm peak}-E_{\rm iso}-t_{\rm jet}$}
\def\sw{{\it Swift}}
\def\sax{{\it Beppo}SAX}

\begin{document}
   

\title{Confirming the $\gamma$--ray burst spectral--energy correlations in the era 
of multiple time breaks}

\author{G. Ghirlanda$^1$, L. Nava$^{1,2}$, G. Ghisellini$^1$, C. Firmani$^{1,3}$}

   \offprints{Giancarlo Ghirlanda \\giancarlo.ghirlanda@brera.inaf.it}

\institute{
Osservatorio Astronomico di Brera, via Bianchi 46, I--20387, Merate, Italy  \and 
Univ. degli Studi dell'Insubria, via Valleggio 11, I--22100, Como, Italy. \and
Instituto de Astronom\'{\i}a, U.N.A.M., A.P. 70-264, 04510, M\'exico, D.F., M\'exico
             }

\date{Received ...  / Accepted ...}

\titlerunning{Spectral--energy correlations in the Swift era}

\authorrunning{G. Ghirlanda et al.} 

\abstract{We test the spectral--energy correlation including the new bursts
  detected (mostly) by \sw\ with firm measurements of their redshifts and peak
  energy. The problem of identifying the jet breaks is discussed in the
  complex and multibreak/flaring X--ray light curves observed by \sw.  We use
  the optical data as the most reliable source for the identification of the
  jet break, since the X--ray flux may be produced by a mechanism different
  from the external shocks between the fireball and the circumburst medium,
  which are responsible for the optical afterglow. We show that the presence
  of an underlying SN event in XRF 050416A requires a break to occur in the
  afterglow optical light curve at around the expected jet break time. The
  possible presence of a jet break in the optical light curve of GRB~050401 is
  also discussed. We point out that, for measuring the jet break, it is
  mandatory that the optical light curve extends after the epoch where the jet
  break is expected.  The interpretation of the early optical breaks in
  GRB~050922C and GRB~060206 as jet breaks is controversial because they might
  instead correspond to the flat--to--steep decay transition common in the
  early X--ray light curves.  All the 16 bursts coming from \sw\ are
  consistent with the \ghi\ and \lz\ correlation.  No outlier is found to
  date. Moreover, the small dispersion of the \ghi\ and \lz\ correlation,
  confirmed also by the \sw\ bursts, strengthens the case of using GRBs as
  standard candles.  \keywords{Gamma rays: bursts --- Radiation mechanisms:
    non-thermal --- X--rays: general } } \maketitle

\section{Introduction}

Since the launch of the \sw\ satellite (Gehrels et al. 2004) we are witnessing
the discovery of new properties of Gamma Ray Bursts (GRBs), 
thanks to the ``prompt'' follow up of the GRB emission from the 
X--ray to the optical and NIR band (see e.g. Zhang 2007 
for a recent review).

In the pre--\sw\ era several correlations involving the prompt and afterglow
properties of long Gamma--Ray Bursts with measured redshifts were reported.
Amati et al. (2002), with 12 GRBs detected by \sax\, found that the prompt
emission (rest frame) spectral peak energy (i.e. the peak of the $\nu F_{\nu}$
spectrum) is correlated with the isotropic energy released during the prompt
phase ($E_{\rm peak}\propto E_{\rm iso}^{1/2}$ - so called ``Amati''
correlation). The inclusion of more bursts detected by BATSE(CGRO), Integral and
Hete--II (Ghirlanda, Ghisellini \& Lazzati 2004 - GGL04 hereafter, Lamb et al.
2004, Nava et al. 2006 - N06 hereafter) confirmed this correlation. Recently,
Amati (2006) confirmed his correlation by adding the 17 bursts, detected in the
\sw\ era (up to GRB~061007), with firm measurements of their $E_{\rm peak}$
and redshift $z$.

The possible jetted nature of GRBs implies that, when the fireball is
decelerated by the circumburst medium up to $\Gamma\sim 1/\theta_{\rm jet}$
(where $\theta_{\rm jet}$ is the jet opening angle), an achromatic break
should appear in the GRB afterglow light curve at a characteristic time
$t_{\rm jet}$ (e.g.  Rhoads 1997; Sari, Piran \& Halpern 1999).  By correcting
the GRB prompt emission energy for the collimation factor
($f=1-\cos\theta_{\rm jet}$), Frail et al. (2001, see also Bloom et al. 2003)
showed that the large dispersion of $E_{\rm iso}$ is reduced.  Then GGL04
discovered (with a sample of 15 GRBs with jet break times measured from the
optical light curves) a tight correlation (so called ``Ghirlanda''
correlation) between the GRB peak energies and the collimation corrected
energy, i.e.  $E_{\rm peak}\propto E_{\gamma}^{0.7}$ 
(where $E_{\gamma}=E_{\rm iso}\cdot f$) if the circumburst medium has a constant density 
(homogeneous
medium -- HM) and the radiative efficiency of the prompt phase is the same for
all bursts (a value of 20\% was used -- e.g. Frail et al. 2001).  In the case
of a stratified circumburst density (e.g. $\rho\propto r^{-2}$ -- wind medium, WM), 
N06 showed that this correlation (computed with the 17 GRBs of the
pre--\sw\ era plus the \sw\ burst GRB 050525A) becomes linear: 
$E_{\rm peak}\propto E_{\gamma}$.  Furthermore, Liang \& Zhang (2005) discovered a
completely phenomenological correlation (so called ``Liang--Zhang''
correlation) involving the three observables $E_{\rm peak}$, $E_{\rm iso}$ and
$t_{\rm jet}$ (all quantities computed in the source frame and $t_{\rm jet}$
estimated from the optical light curve) whose consistency with the model
dependent \ghi\ correlations has been demonstrated in N06.

The quantities needed to test the \ghi\ and \lz\ correlations are the redshift
$z$, the peak energy $E_{\rm peak}$ and the jet break time $t_{\rm jet}$.
Despite $\sim 50$ (i.e. about $\sim 1/4$) of the \sw\ GRBs have a measured
redshift, both the \ghi\ and the \lz\ correlations were hardly tested with new
bursts due to: (i) the difficulty to measure the peak energy $E_{\rm peak}$
with the relatively small energy band (15--150 keV) of the BAT instrument
on--board \sw; and (ii) the unexpected (and unforeseen in the pre-\sw\ era)
complexity of the early ($\le 1$d) afterglow light curves (Nousek et al. 2006;
O'Brien et al. 2006; Borrows et al. 2005).

A typical \sw\ X--ray light curve, in fact, presents at least three phases
(Nousek et al. 2006): (a) a steep initial decay followed by (b) a flat phase
and finally (c) by a steeper decay (similar to what observed in the the
pre--\sw\ afterglow light curves).  These phases are separated by
corresponding breaks at $\sim 500$ s [from (a) to (b)] and between 10$^3$ and
10$^4$ s [from (b) to (c)].  Also, an intense (long--lasting) flaring activity
is superposed to the typical bare power law decay of the afterglow X--ray
light curve (Burrows et al. 2007).  The optical light curve tracks the X--ray
light curve in some cases, but more often it is different.  Also in the
optical light curve there can be a multiplicity of breaks, also not
simultaneous with those of the X--ray light curve (e.g. Panaitescu 2007).

In a recent paper, Willingale et al. (2007, W07 hereafter), studied the \sw\ 
GRB X--ray emission and suggested that both the very early and the late (up
to 10$^{5}$--10$^{6}$ s) emission can be reproduced with a model having the 
same functional form, i.e. an exponential smoothly connected with a power law.  
For the afterglow phase, this functional form allows to define the characteristic 
time $T_a$ where the power law sets in and dominates the emission, which W07 
estimated for a large number of \sw\ bursts.
Intriguingly, they found that if $T_a$ is treated as a jet break time
(i.e. with the same formalism which is used to derive the collimation angle
$\theta_{\rm jet}$ from the measure of $t_{\rm jet}$), then the collimation
corrected energetics correlates strongly with $E_{\rm peak}$, for those \sw\ 
bursts of known $E_{\rm peak}$ and $z$.  

Trying to explain the different optical and X--ray behavior, Uhm \&
Beloborodov (2007) and Grenet, Daigne \& Mochkovitch (2007) have proposed that
the X--ray flux may be dominated by the reverse shock emission in slow shells,
while the optical flux is instead due to the standard forward external shock.
In this case only the optical light curve (and even not always, if the reverse
shock contributes also in the optical) can show a jet break time. Another
possibility is that the X--ray flux is dominated by a late activity of the
central engine (late prompt emission model, see Ghisellini et al. 2007).

{\em This complexity requires care when identifying a break with the jet break
  time.}  Since the pre--\sw\ \ghi\ and \lz\ relations made use of break times
found in the optical light curves (typically starting several hours after the
GRB trigger), it is safer to use {\it only} the optical light curves to find
$t_{\rm jet}$, and relax the requirement that the break should be present also
in the X--ray light curve which, as mentioned above, could be produced by a
different mechanism.  In addition, an early optical break should be tested for
being the jet break: if the light curve does not extend to (and even after)
the epoch when the jet break is expected (according to the \ghi\ correlation),
this early break cannot be claimed to be the jet break.  This is, indeed, the
case of GRB~050922C and 060206 (discussed below).

The relevance of the \ghi\ (and of its phenomenological form, i.e. the \lz\ 
correlation) is twofold: (1) they represent a new tool to understand the GRB
physics and some interpretations have already been proposed (see e.g.  Eichler
\& Levinson 2005; Thompson 2006; Thompson, Meszaros \& Rees 2007); (2) the
tightness of these correlations makes them a new tool to standardize the GRB
energetics for a cosmological use (e.g. Ghirlanda, Ghisellini \& Firmani
2006 for a review). More recently, Firmani et al. (2006 - F06 hereafter)
discovered a new correlation which is based only on prompt emission properties
and is a new tool to use GRBs as standard candle which does not require the
use of the jet break time (Firmani et al. 2006a, 2006b).

In this paper we update the \ghi\ and \lz\ correlation by including all the
possible GRBs detected in the \sw\ era.  With a sample which is doubled with
respect to the original one (GGL04) we find no new outlier with respect
  to these correlations, despite previous claims of the contrary (Sato et al.
  2006 - S06 hereafter; Mundell et al. 2007).  We will then discuss these
  cases in more detail.  
  
  In Sec. 2 we present and discuss the sample of GRBs, in Sec. 3 we find
  (GRB~050416A and GRB~050401) and discuss (GRB~050922C and GRB~060206) some
  jet break times derived in the optical, while in Sec. 4 we present the
  results of the statistical analysis of the \ghi\ and \lz\ correlation. In
  Sec. 5 we discuss our results.
  
  We use a standard cosmology with $\Omega_{\rm M}=0.3$ and
  $\Omega_\Lambda=h=0.7$.

\section{The sample}

Before the \sw\ mission the jet break time $t_{\rm jet}$ was measured from the
optical light curves of GRB afterglows.
Since it should be achromatic, we were hoping that the \sw\ capability to
localize and promptly follow the X--ray emission of the burst would greatly
help to easily measure several $t_{\rm jet}$ from the X--ray light curve.
Indeed, \sw\ has observed several X--ray afterglows starting $\sim 100$
seconds after the GRB onset. In a few cases the X--ray emission was measured
also simultaneously with the prompt $\gamma$--ray emission (e.g. Butler \&
Kochevski 2007).  But, unexpectedly, these light curves are complex
(steep--flat--steep structure with multiple breaks and several superposed
early and late time flares).  As a consequence, {\it the jet break can be hardly
  identified using the X--ray data.}  It is therefore reasonable to rely on
the optical light curves to search for possible jet breaks. Also in this case,
however, we should keep in mind that the optical emission often shows a
complexity similar to what seen in the X--rays.

After having selected all \sw\ GRBs with measured redshift, we have searched
for any published information on their prompt emission spectral properties in
order to obtain $E_{\rm peak}$ and to estimate $E_{\rm iso}$.  Finally, we
searched for evidence of jet breaks in the optical light curves.  In several
cases we could only set a lower limit on $t_{\rm jet}$ although the optical
emission was fairly well sampled. For the reasons discussed above we did not
include in our sample the GRBs for which a jet break was claimed only in the
X--ray light curve.  We present our sample in Tab.~1 dividing the pre--\sw\ 
GRBs from those identified in the \sw\ era and added in this paper (from
GRB~050525A to GRB~061121).

The typical time--averaged spectrum of GRBs (e.g. Preece et al. 2000; Kaneko
et al. 2006) is best fitted by a smoothly joint double power law model (the
Band model -- Band et al. 1993). However, for some bursts detected by Hete--II
and included in the pre--\sw\ sample of N06 and F06, the time--averaged
spectrum reported in the literature was fitted with a
power law ending with a high--energy exponential cutoff (CPL model).  The isotropic
energy used to define the Amati correlation as well as the \ghi\ and \lz\ 
correlations is derived by extrapolating the best fit spectral model in the
rest frame 1--10$^4$ keV energy band (e.g. see Eq.~1 in GGL04). As a
consequence those bursts whose spectrum is fitted with a CPL model
underestimate the value of $E_{\rm iso}$ (see also F06). The same happens for
most of the \sw--BAT bursts whose spectra, due to the limited energy range of
this instrument, are typically fitted with a CPL model.

In all these cases (i.e. GRB~011211, 020813, 021004, 030226, 030429, 041006,
050318, 050525A, 050820A 050922C, 051109A, 060124, 060206, 060418, 060927) we
estimated $E_{\rm iso}$ as the (logarithmic) average between the value
obtained assuming the cutoff--power law model that was reported in the
literature
and the value obtained with the same spectral parameters but with a Band model
having the high energy spectral component fixed, i.e $\beta=-2.3$ (photon
spectral index) which is the average value found from the spectral analysis of
GRBs (e.g. Kaneko et al. 2006).  In these cases we adopted an error on
$E_{\rm iso}$ which accounts for these two extreme values.

Note that, among the pre--\sw\ bursts, we have updated the values of 
$E_{\rm iso}$ with respect to the sample of N06 and F06 (for GRB~011211, 020813,
021004, 030226, 030429, 041006). Moreover, we have updated the jet break time
of GRB~030226 with the most recently published estimate (Klose et al. 2004)
and we have corrected a typo in the N06 table on the rest frame value of
$E_{\rm peak}$ for GRB~030328 and GRB~050525A. For
GRB~061121 and GRB~061007 we averaged the $E_{\rm peak}$ and $E_{\rm iso}$
values found by the analysis of the Konus--Wind and RHESSI spectra
(Golenetskii et al., 2006c; Golenetskii et al., 2006b; Bellm et al. 2006).
For GRB~050416A we adopted the spectral parameters (and $E_{\rm iso}$)
calculated by S06.

Finally, in a recent work, Schaefer (2006) used some other bursts with
published $z$ and $E_{\rm peak}$ which we do not include in our sample.  The
breaks of GRB~050318, 050505, 051022 and 060210 were in fact identified in the
X--ray light curve, so we do not include these cases for the reasons explained
above.  However, Schaefer (2006), by adopting the jet breaks observed in the
X--ray light curves of the bursts above, found that they are consistent with
the \ghi\ correlation. Only GRB~060526 (present in the sample of Schaefer
2006) is included in our sample because Dai et al.  (2006) have shown the
presence of an achromatic optical/X--ray jet break at 2.77
d\footnote{GRB~060526 is present also in the sample of Schaefer (2006) but he
  adopted the early jet break observed in the X--ray light curve (Moretti et
  al. 2006).}.  We also considered the most distant GRB~050904 (also included
in Schaefer 2006) at $z=6.29$ which has an achromatic jet break (Tagliaferri
et al. 2006). We did not include this burst in our sample because its $E_{\rm
  peak}$ estimate is still uncertain: Amati (2006) lists a lower limit of 1100
keV (rest frame) while Schaefer (2006) lists a value of 3178 keV (rest frame).
Using the latter value and the density inferred in the homogeneous case
($n=680$ cm$^{-3}$ -- Frail et al. 2006), this burst is consistent with the
\ghi\ correlation.

In this search we found two more bursts (i.e. GRB 060115 and GRB 060707,
listed by Amati 2006) which, having a firm estimate of $E_{\rm peak}$ and $z$,
are candidates to be included in the present sample. However, for these bursts
the optical light curves (publicly available at present only in form of GCNs)
are poorly sampled\footnote{see also http://grad40.as.utexas.edu/grblog.php}.
Note finally that we did not include in the present sample GRB~060218
(associated with SN2006aj, see e.g.  Mazzali et al. 2006; Campana et al. 2006)
because the SN dominates the optical emission, making impossible to estimate
$t_{\rm jet}$.  It is consistent with the $E_{\rm peak}$--$E_{\rm iso}$
correlation (Amati et al. 2007), and its possible jet opening angle is
discussed in Ghisellini et al. (2006).

\begin{table*}
\tabcolsep=4.5pt
\begin{center}
\begin{tabular}{llllllllll}
\\
\hline
\hline
\\
GRB  & $z$ & $t_{\rm jet}$ &  $E_{\rm peak}$ &  ref$^a$ &$E_{\rm \gamma, iso}$  & $\theta_{\rm j}$ & $E_{\gamma}$ & $\theta_{\rm j,w}$ & $E_{\gamma,\rm w}$ \\ 
     &     &  days         &  keV               &  & erg                    &
     deg                 & erg   & deg & erg \\
\\
\hline
970828  &0.958 &2.2  (0.4)  &583 (117)  & &2.96e53 (0.35) &5.91 (0.80)  &1.57e51 (0.46)  & 3.40 (0.18)  & 5.21e50 (0.84)  \\ 
980703  &0.966 &3.4  (0.5)  &499 (100)  &  &6.90e52 (0.82) &11.02 (0.80) &1.27e51 (0.24)  & 5.45 (0.26)  & 3.12e50 (0.47)  \\ 
990123  &1.600 &2.04 (0.46) &2031 (161) & &2.39e54 (0.28) &3.98 (0.57)  &5.76e51 (1.78)  & 1.84 (0.12)  & 1.24e51 (0.21)  \\ 
990510  &1.619 &1.6  (0.2)  &422  (42)  & &1.78e53 (0.19) &3.74 (0.28)  &3.80e50 (0.70)  & 3.31 (0.14)  & 2.98e50 (0.40)  \\ 
990705  &0.843 &1.0  (0.2)  &348  (28)  &  &1.82e53 (0.23) &4.78 (0.66)  &6.33e50 (1.92)  & 3.20 (0.19)  & 2.84e50 (0.49)  \\ 
990712  &0.433 &1.6  (0.2)  & 93  (16)  &  &6.72e51 (1.29) &9.46 (1.20)  &9.15e49 (2.90)  & 8.74 (0.50)  & 7.81e49 (1.74)  \\ 
991216  &1.02  &1.2  (0.4)  &642 (129)  &  &6.75e53 (0.81) &4.44 (0.70)  &2.03e51 (0.68)  & 2.36 (0.21)  & 5.72e50 (1.22)  \\ 
011211  &2.140 &1.56 (0.16) &185  (25)  &  &6.64e52 (1.32) &5.25 (0.65)  &2.78e50 (0.88)  & 4.03 (0.23)  & 1.64e50 (0.37)  \\ 
020124  &3.198 &3.0  (0.4)  &390 (113)  &  &2.15e53 (0.73) &5.19 (0.69)  &8.82e50 (3.80)  & 3.29 (0.30)  & 3.54e50 (1.36)  \\ 
020405  &0.695 &1.67 (0.52) &617 (171)  &  &1.25e53 (0.13) &6.27 (1.03)  &7.47e50 (2.57)  & 4.08 (0.33)  & 3.17e50 (0.62)  \\ 
020813  &1.255 &0.43 (0.06) &478  (95)  &  &6.77e53 (1.00) &2.74 (0.35)  &7.74e50 (2.28)  & 1.77 (0.10)  & 3.24e50 (0.58)  \\ 
021004  &2.335 &4.74 (0.5)  &267 (117)  &  &4.09e52 (0.71) &8.27 (1.02)  &4.25e50 (1.28)  & 5.91 (0.30)  & 2.18e50 (0.44)  \\
030226  &1.986 &0.84 (0.10) &290  (63)  &  &6.7e52  (1.2)  &4.23 (0.53)  &1.83e50 (0.56)  & 3.49 (0.19)  & 1.24e50 (0.26)  \\ 
030328  &1.520 &0.8  (0.1)  &328  (35)  &  &3.61e53 (0.40) &3.59 (0.45)  &7.08e50 (1.93)  & 2.36 (0.10)  & 3.06e50 (0.42)  \\ 
030329  &0.169 &0.5  (0.1)  &79   (3)   &  &1.66e52 (0.20) &5.67 (0.50)  &8.13e49 (1.75)  & 5.49 (0.32)  & 7.60e49 (1.28)  \\ 
030429  &2.656 &1.77 (1.0)  &128  (37)  &  &1.73e52 (0.31) &6.15 (1.49)  &9.95e49 (5.13)  & 5.60 (0.83)  & 8.26e49 (2.86)   \\ 
041006  &0.716 &0.16 (0.04) &108  (22)  &  &8.3e52 (1.3)   &2.72 (0.41)  &9.38e49 (3.17)  & 2.51 (0.18)  & 7.94e49 (1.71)   \\
\hline
050318  &1.44  &$>$0.26 (0.13)&115 (27) &[1]1  &2.00e52 (0.31) &$>$3.42 (0.76) &$>$3.57e49 (1.67)& $>$3.70 (0.48)  & $>$4.17e49 (1.27) \\
050401  &2.9   &1.5     (0.5) &501 (117)&[2]2 &4.1e53 (0.8)   &   3.8 (0.65) &9.0e50 (3.5)   & 2.40    (0.23)  &     3.59e50  (0.98) \\
050416A &0.653 &1.0   (0.7) &28.6 (8.3) &[3]3 &8.3e50 (2.9)   &   9.77 (2.83) &1.2e49 (0.8)   & 12.66 (2.47)    &    2.02e49 (1.05) \\
050525A &0.606 &0.3  (0.1)  &127  (5.5)  &[4]4 &2.89e52 (0.57) &4.03 (0.69)   &7.16e49 (2.83)    & 3.88 (0.38)   &6.63e49 (1.83)  \\
050603  &2.821 &$>$2.5 (1.25)&1333 (107)&[5]5 &5.98e53 (0.4)  &$>$4.42 (0.97) &$>$1.78e51 (0.79)& $>$2.49 (0.31)  & $>$5.65e50 (1.47) \\
050820A &2.612 &15.2 (8)    &1325 (277) &[6]6 &9.75e53 (0.77) &6.65 (1.53)   &6.55e51 (3.05)    & 3.50 (0.47)   &1.82e51 (0.51)  \\
050922C &2.198 &$>$1.2 (0.6)&417 (118)  &[7]7 &4.53e52 (0.78)&$>$4.95 (1.09)&$>$1.69e50 (0.80)  & $>$4.13 (0.55) &$>$1.18e50 (0.37)\\
051109A &2.346 &$>$0.64 (0.32)&539 (381)&[8]8 &7.52e52 (0.88) &$>$3.61 (0.80) &$>$1.49e50 (0.68)& $>$3.08 (0.39)  & $>$1.08e50 (0.31) \\
060124  &2.297 &1.1  (0.1) &636  (162) &[9]9 &4.3e53 (0.34)   &3.61 (0.43)   &8.55e50 (2.15)    &2.30 (0.06)    &3.47e50 (0.34)  \\
060206  &4.048 &$>$2.3 (0.11)&381  (98) &[10]10&4.68e52 (0.71) &$>$5.31 (0.63) &$>$2.00e50 (0.6)  & $>$4.3 (0.2) &$>$1.32e50 (0.23)  \\
060418  &1.489 &$>$5.0   (2.5) &572 (114)&[11]11&1.28e53 (0.10) &$>$8.16 (1.80) &$>$1.30e51 (0.58)& $>$4.85 (0.61)  & $>$4.58e50 (1.21) \\
060526  &3.21  &2.77   (0.30) &105 (21) &[12]12&2.58e52 (0.3)  &  6.56  (0.80) &1.70e50 (0.5)    & 5.47 (0.20)     & 1.18e50 (0.15)    \\
060614  &0.125 &1.38 (0.04) &55   (45)  &[13]13 &2.5e51 (1.0)   &11.12 (1.4)   &4.71e49 (2.22)    &11.47 (1.15)   &5.01e49 (2.23)  \\
060927  &5.6 &$>$0.16 (0.08) &473 (116) &[14]14&9.55e52 (1.48) &$>$1.62 (0.36) &$>$3.79e49 (1.78)& $>$1.73 (0.23)  & $>$4.35e49 (1.32)  \\
061007  &1.261 &$>$1.74 (0.87) &902 (43)&[15]15  &8.82e53 (0.98) &$>$4.47 (1.00) &$>$2.69e51 (1.22)& $>$2.35 (0.30)  & $>$7.4e50 (2.1)   \\
061121  &1.314 &$>$3.87 (1.9) &1289 (153)&[16]16 &2.61e53 (0.3)  &$>$6.97 (1.52) &$>$1.93e51 (0.87)& $>$3.87 (0.49)  & $>$5.96e50 (1.65) \\
\hline 
\hline
\\
\end{tabular}
\end{center}
\caption{ 
Burst with firm estimates of the redshift and of the spectral peak energy 
$E_{\rm peak}$ (rest frame). 
Bursts of the pre--\sw\ era (from GRB~970828 to GRB~041006) and bursts of the 
\sw\ era (from GRB~050525A) added in this work are separated by the horizontal line.  
For each burst we report the value of the jet opening angle $\theta_{\rm j}$ 
($\theta_{\rm j, w}$) and of the collimation corrected energy $E_{\gamma}$ 
($E_{\gamma, \rm w}$) computed in the case of a homogeneous (wind) external medium. 
For  those bursts with only a lower limit on $t_{\rm jet}$ we adopted a 50\% uncertainty on 
its value to compute the error on the collimation corrected energy. 
$a$ References in square parenthesis are for the jet break time:
[1] Still et al., 2005
[2] This paper;
[3] This paper;
[4] Blustin et al., 2005; Mirabal et al. 2005; Della Valle et al., 2006a;
[5] Grupe et al., 2006; 
[6] Cenko et al., 2006;
[7] This paper; Covino et al., 2005;
[8] Pavlenko et al., 2005; http://grad40.as.utexas.edu/grblog.php;
[9] Romano P. et al., 2006; Curran et al. 2006;
[10] This paper; Stanek et al., 2006;
[11] Karimov et al., 2005; Vergani et al. 2007;
[12] Dai et al., 2006;
[13] Della Valle M. et al., 2006b;
[14] Atoniuk et al. 2006; http://grad40.as.utexas.edu/grblog.php;
[15] Shady et al., 2006;
[16] Halpern J. P. \& Armstrong E., 2006.
Following references are for $E_{\rm peak}$ and the spectral properties (in
case of assimetric 
errors on $E_{\rm peak}$  we computed the logarithmic average value): 
(1) Perri et al., 2005; 
(2) Golenetskii et al., 2005a (we computed the wighted average of the values
reported in this reference); 
(3) Sato et al., 2006; 
(4) Blustin et al., 2006; 
(5) Golenetskii et al., 2005b; 
(6) Cenko et al., 2006; 
(7) Crew et al., 2005; 
(8) Golenetskii et al., 2005c; 
(9) Romano et al., 2006; 
(10) Palmer et al., 2006; 
(11) Golenetskii et al., 2006a (20\% error on $E_{\rm peak}$ is assumed);
(12) Schaefer 2006; 
(13) Amati 2006;
(14) Stamatikos et al., 2006;
(15) Golenetskii et al., 2006c;
(16) Golenetskii et al., 2006b; Bellm et al., 2006 (we computed the wighted average of the values
reported in these references).
}  
\label{tabout}
\end{table*}

\section{Optical breaks}

In order to rely on published data we have considered all bursts with a
published optical light curve and we have checked for the presence of jet
breaks. In two cases (GRB 050401 and GRB 050416A), discussed below, we present our
estimate of the jet break time from the fitting of the available optical data.
Sato et al. 2006 claimed that these two bursts are outliers for the \ghi\ 
correlation. We show that, instead, our estimate of the jet break time
makes them consistent with this correlation.

\begin{figure}
\centerline{\psfig{figure=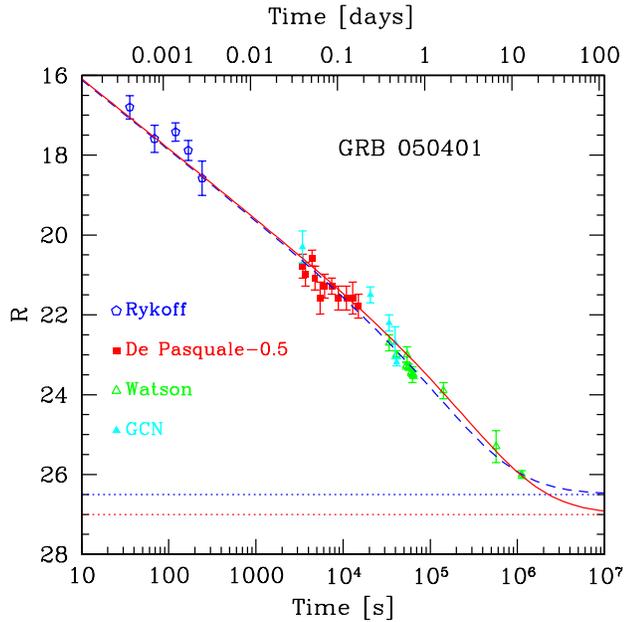,width=9.5cm,height=9.5cm}}
\vskip -0.5 true cm
\caption{The optical light curve of GRB 050401.
Data from: D'Avanzo et al. (2005); De Pasquale et al. (2006); 
Greco et al. (2005); Kahharov et al. (2005); McNaught et al. (2005);
Misra et al. (2005); Rykoff et al. (2006); Watson et al. (2006).
The dashed and solid lines are two fits with break time set
at 1 and 2 days, respectively. 
The dotted lines correspond to the magnitude assumed for the host galaxy.
}
\label{lc050401}
\end{figure}
\begin{figure}
\psfig{figure=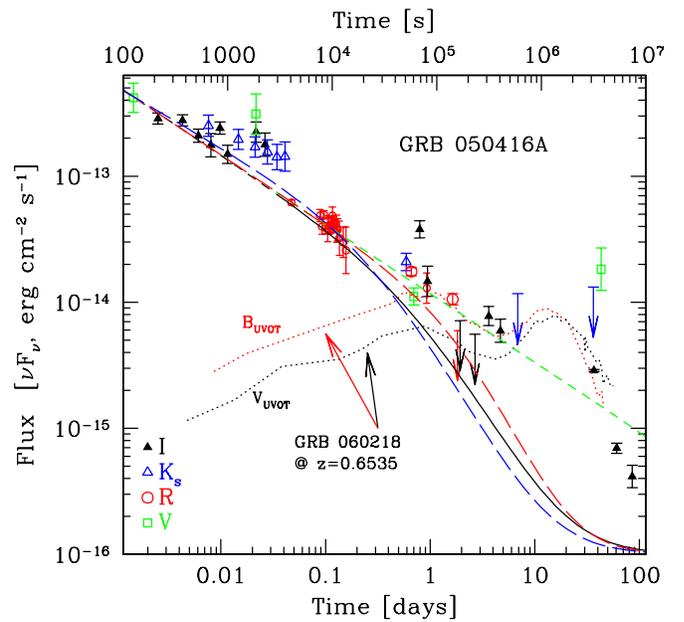,width=9.5cm,height=9.5cm}
\vskip -0.5 true cm
\caption{The optical and near  IR light curve of GRB 050416A.
  Data from Soderberg et al. (2006) and Holland et al. (2007).  These data
  have been fitted with a smoothly broken power law plus a constant (see Eq.
  \ref{lc}), assuming three different break times: 0.3 and 2 days (long dashed
  lines) and 0.8 day (solid line).  The short dashed line is an example of
  unbroken power law, which clearly exceeds late time data.  In this burst the
  rebrigthening at $\sim$10 days is most likely due to an underlying
  supernova.  The flux associated to the supernova is likely to contribute to
  the the light curve even for earlier times, as shown by the light curve of
  GRB 060218 (from Campana et al. 2006, dotted lines), assuming to lie at the
  same redshift of GRB 050416A ($z=0.6535$).  }
\label{lc050416}
\end{figure}

\subsection{GRB 050401}

We have collected the photometric data for this bursts from the literature,
including the $R=26\pm0.1$ magnitude Subaru observation reported in Watson et
al. (2006).  In that paper, it is not specified if this magnitude is inclusive
or not of the host galaxy.  We have assumed this magnitude corresponds to the
sum of the afterglow plus the host galaxy fluxes.  De Pasquale et al. (2006)
pointed out that the data in their list, while internally consistent, may have
up to 0.5 magnitude uncertainty in absolute normalization when compared with
data of other sources, due to the differences in the used reference stars.  We
have then ``renormalized" these data to have a rough agreement with other
observations made close in time.  In addition, the magnitudes given in De
Pasquale et al. (2006) were de--reddened by galactic extinction.  For
consistency with the other data, we have instead taken the observed
magnitudes.  We have then fitted a broken power law plus a constant to the
data:
\begin{equation}
F(t)\, =\, F_{\rm host} \, +\, 
\frac{F_{\rm b} \, (t/t_{\rm b})^{-\alpha_1}} 
 {1+(t/t_{\rm b})^{\alpha_2-\alpha_1}}
\label{lc}
\end{equation}
The dashed line in Fig. \ref{lc050401} corresponds to 
$t_{\rm b}=1$ d; $R_{\rm host}=26.5$; $\alpha_1=0.7$ and $\alpha_2=1.4$;
while the solid line corresponds to
$t_{\rm b}=2$ d; $R_{\rm host}=27$; $\alpha_1=0.7$ and $\alpha_2=1.4$.

These results do not greatly change by adopting a different renormalization
for the data of De Pasquale et al. (2006).  We conclude that the data show the
existence of a possible break in the optical light curve of GRB 050401, and set
$t_{\rm b}=1.5\pm 0.5$ days.

\subsection{GRB 050416A}

In Fig. \ref{lc050416} we show the optical and near IR light curve of
GRB 050416A, with data taken from Soderberg et al. (2006) and Holland 
et al. (2007).
Since the rebrightening at $\sim$10 days is most likely
due to an underlying supernova (Soderberg et al. 2006),
in Fig. \ref{lc050416} we also show the light curve 
of GRB 060218 (from Campana et al. 2006, dotted lines),
assuming to lie at the same redshift of GRB 050416A ($z=0.6535$).
It can be seen that its flux can contribute to the light curve
of the afterglow even much earlier than 10--20 days.
Note that the rest frame wavelengths of the $R$ and $I$ 
filters for GRB 0504016A roughly corresponds to
the rest frame wavelengths of the $B$ and $V$
bands for GRB 060218 ($z=0.033$). 

Using Eq. \ref{lc}, we have fitted these data assuming three different break
times: 0.3 and 2 days (long dashed lines) and 0.8 day (solid line).  The
values of the decay indices changes little in all cases, being in the ranges
$0.48<\alpha_1<0.55$ and $1.5<\alpha_1<1.8$.  Note that an unbroken power law
clearly exceeds late time data (short--dash line).

We conclude that the data require the light curve to break,
but the uncertainties due to the paucity of data and to
the contribution from the supernova allow
to determine $t_{\rm b}$ only with a large error,
therefore we set $t_{\rm b}=1.0\pm 0.7$ days.

\subsection{GRB 050922C \& GRB 060206} 

These two bursts have an optical light curve showing early breaks at
0.11$\pm$0.03 (Li et al. 2005) for GRB~050922C and $\sim0.6$ 
days (Monfardini et al. 2006, Stanek et al. 2006) for GRB~060206. 
The jet breaks predicted by
the \ghi\ correlation are instead $t_{\rm jet}=8.0^{+5.7}_{-3.3}$ days for
GRB~050922C and $t_{\rm jet}=10.3^{+7.4}_{-4.3}$ days for GRB~060206 (in the
HM case -- to be consistent with the \ghi\ correlation within its 1$\sigma$
scatter) and $t_{\rm jet}=8.7^{+6.8}_{-3.8}$ for GRB~050922C and 
$t_{\rm jet}=11.2^{+8.7}_{-4.9}$ for GRB~060206 (in the WM case). 
The available
optical light curve, in both cases, does not extend to such epochs and
therefore it is not possible to say if they really have a jet break when it is
expected according to the \ghi\ correlation.
  
  For this reason these two bursts cannot be considered outliers of the
  \ghi\ correlation (as they would be, adopting the early optical break as the 
  jet break) nor
  as lower limits for this correlation (by adopting the latest optical
  observation available).  

\subsubsection{On the possible nature of the optical breaks in GRB 050922C \&
  GRB 060206}

\begin{figure}
  \psfig{figure=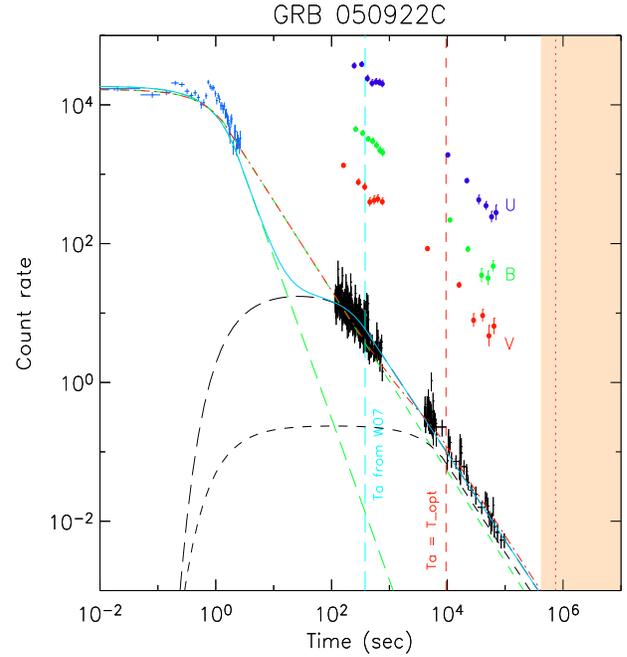,width=9.cm,height=9.cm}
  \vskip -0.2 true cm
\caption{Early BAT and XRT light curve of GRB~050922C obtained from the Butler archive
  http://astro.berkeley.edu/$\sim$nat/swift/. The data points are the count
  rate at 1 keV. In the case of the early light curve these have been obtained
  by extrapolating at this energy the BAT spectrum. The solid line corresponds
  to the double component model of W07 (with the parameters listed in their
  Table 1) with the two components shown by the long--dashed lines. The value
  of $T_{a}=380$ s derived by W07 is shown by the vertical long--dashed line.
  The dot--dashed line show our fit to the data (the two model components are
  represented by the dashed lines) by assuming a flatter early component (i.e.
  $\propto t^{-1.3}$) and $T_{a}=t_{opt}=0.11$ days. The dotted vertical line
  marks where the jet break is expected based on the \ghi\ correlation (in the
  WM case) and the shaded region its $1\sigma$ interval. Optical fluxes (in
  arbitrary units, UVOT/\sw, $U$, $B$, $V$ bands, from top to bottom,
  displaced for clarity, data from Li et al., 2005) are also shown.}
\label{050922C}
\end{figure}

The early optical break in GRB~050922C (at 0.11$\pm$0.03 days after the BAT
trigger) is 25 times larger than the break that W07 derive ($T_a=380$ s) from
a multi--component fit to the BAT--XRT light curve up to $10^5$ s (solid line
in Fig. \ref{050922C}).

With $T_a=380$ s the burst is an outlier with respect to the \ghi\ 
correlation derived by the same authors, which uses $T_a$ as a jet break time.

In Fig. \ref{050922C} we show the early BAT light curve and the XRT light curve
starting 100 s and extending to 1 day after the trigger\footnote{Data are
  taken from the public archive of Butler
  http://astro.berkeley.edu/$\sim$nat/swift/}. 

We tried to ``match'' the X--ray light curve with the functional form proposed
by W07 by assuming $T_{a}$ equal to the break observed in the optical light
curve ($t_{opt}$ = 0.11 days -- Li et al. 2005). This is shown by the
dot--dashed curve in Fig. \ref{050922C} (while the functional form fitted by
W07 is the solid line in the same figure). We note that the two fits are
consistent with the available data and suggest that $T_{a}$ might lie between
the two values, i.e. 380 s (derived by W07) and the break observed in the
optical (at 0.11 days).  With $T_a=t_{opt}\sim 10^4$ s GRB~050922C is not an
outlier with respect to the \ghi\ correlation defined by W07.

The published optical data of GRB~050922C end at $t=1.2$ days (Covino et al.
2005). If, according to the above interpretation, the early optical break is
$T_{a}$, we are authorized to assume $t_{\rm jet}>1.2$ days and treat this as
a lower limit for the \ghi\ correlation.

\begin{figure}
  \psfig{figure=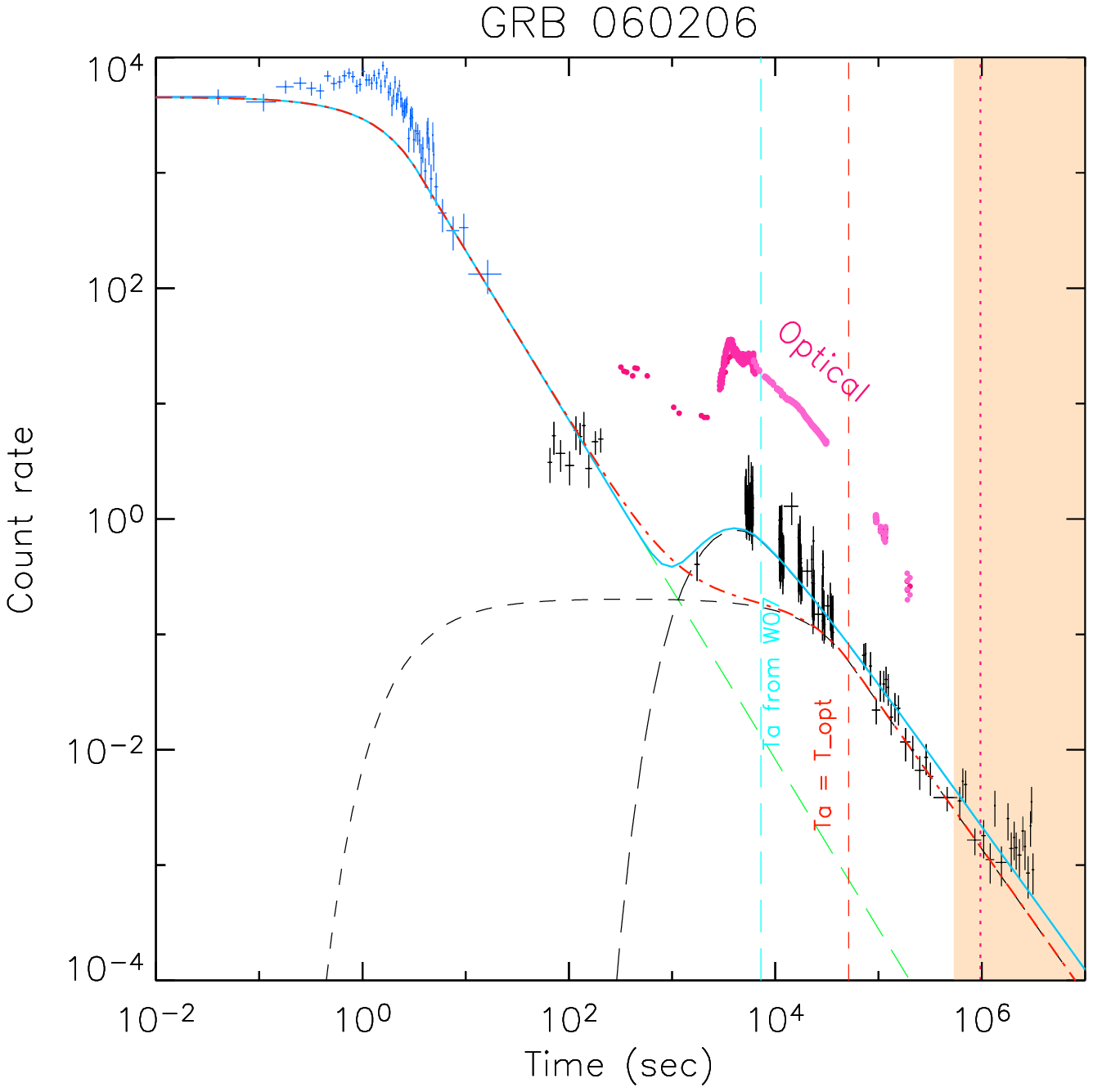,width=9.cm,height=9.cm}
  \vskip -0.2 true cm
\caption{Early BAT and XRT light curve of GRB~060206. The solid line corresponds to
  the double component model of W07 (with the parameters listed in their table
  1) with the two components represented by the long--dashed lines. The value
  of $T_{a}=7244$ s derived by W07 is shown by the vertical long--dashed line.
  The dot-dashed line show our fit to the data (the two model components
  correspond to the dashed lines) by assuming $T_{a}=t_{opt}=0.6$ days and by
  excluding the X--ray flare at 10$^4$ s. The optical R--band data are shown
  (Monfardini et al.  2006; Stanek et al. 2006; Wozniak et al. 2006). The
  dotted vertical line marks where the jet break is expected, based on the
  \ghi\ correlation (in the WM case) and the shaded region its $1\sigma$
  interval.}
\label{060206}
\end{figure}

For GRB~060206 the observed early optical break (Monfardini et al. 2006,
Stanek et al. 2006) in only 10 times larger than the $T_{a}$ derived by W07.
However, as shown by Monfardini et al. (2006) and Stanek et al. (2006), the
optical light curve presents a strong flare peaking at 3 -- 4$\times 10^{3}$ s.
The same flaring structure is present in the X--ray light curve.

We show the X--ray and optical light curve of GRB~060206 in Fig. \ref{060206}
and the model function of W07 (solid line) which gives $T_{a}\sim 7000$ s.
Similarly to the case of GRB~050922C we fitted the X--ray data by assuming
$T_{a}=t_{opt}=0.6$ days (dot-dashed line in Fig. \ref{060206}) but
excluding the X--ray flare coincident with that observed in the optical. 

The optical data of GRB~060206 (Stanek et al. 2006) end at 2.33 days,
i.e. before the epoch at which the jet break is predicted by the \ghi\
correlation (shaded region in Fig. \ref{060206}). Therefore we assume this
value as a lower limit on the jet break time of this burst.

\begin{figure*}
\psfig{figure=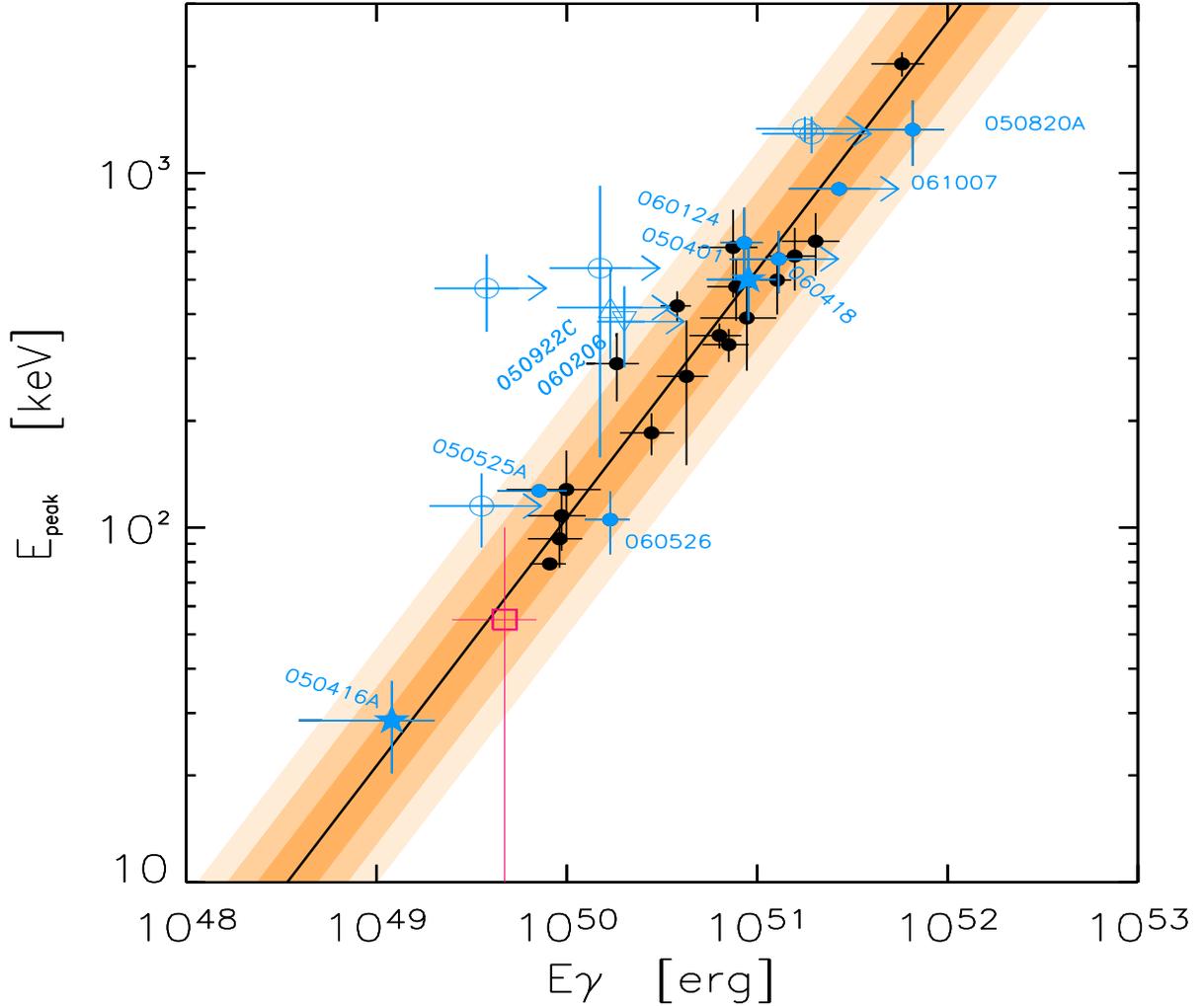,width=18cm,height=15cm}
\vskip -0.5 true cm
\caption{The $E_{\rm peak}$--$E_{\gamma}$ correlation with the inclusion 
  of 16 \sw\ GRBs for a homogeneous circumburst medium case.  The best fit
  correlation (solid line) and its 1, 2, and 3 $\sigma$ scatter (solid filled
  regions) are reported.  The names of the 8 \sw\ bursts used in the fit are
  reported.  The open triangles and open circles are the burst not included in
  the fit because only a lower limit on their jet break was found.  The open
  square corresponds to GRB 060614 which is not included in the fit.
  GRB~050401 and GRB~050416A whose jet break time was estimated in this work
  are shown with filled star symbols.  Note that we did not show, in this
  figure, GRB 980425 and GRB 031203 which are outliers not only of the $E_{\rm
    peak}$--$E_{\gamma}$ correlations, but also of the Amati correlation (but
  see Ghisellini et al. 2006).  GRB 060218, instead, obeys the Amati
  correlations, but the presence of the SN lightcurve made any estimate of
  $t_{\rm jet}$ impossible.  }
\label{ghirla_homo}
\end{figure*}

\subsection{GRB 061007}

Mundell et al. (2007) claim that GRB~061007 is an outlier with respect to the
\ghi\ and \lz\ correlations because no break is observed up to 10$^6$ s in the
X--ray light curve.  For the arguments detailed above (see also Sec. 5) we do
not consider the X--ray light curve. We instead take the lower limit on the
jet break set by the latest optical observation at 1.7 d (Mundell et al. 2007;
Shady et al.  2006). In this case the \ghi\ correlation predicts a jet break
between 0.7 and 2.2 days (1$\sigma$ in the HM case) and 0.8 and 2.4 days
(1$\sigma$ in the WM case).

\begin{figure*}
\psfig{figure=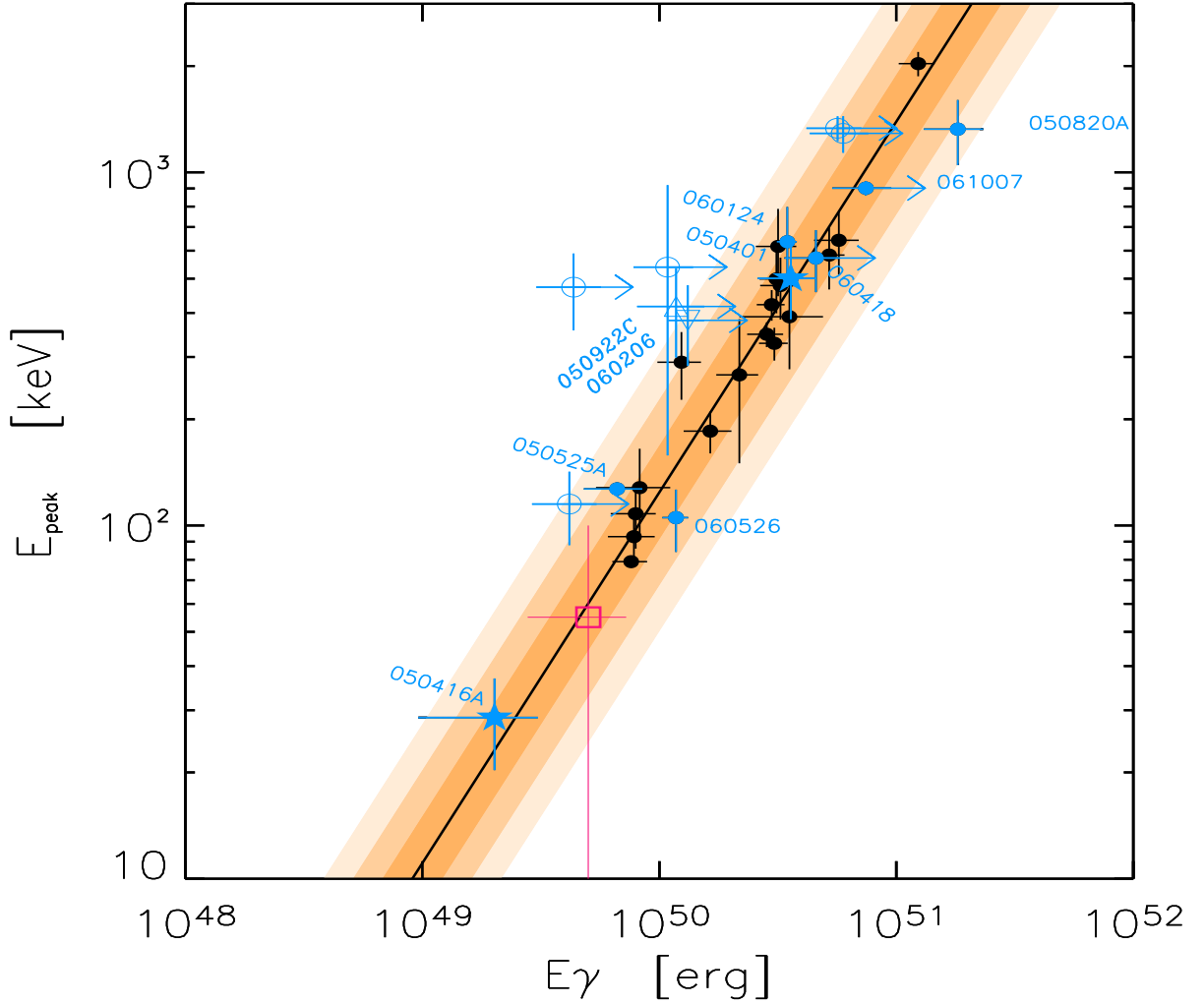,width=18cm,height=15cm}
\vskip -0.5 true cm
\caption{
The $E_{\rm peak}$--$E_{\gamma,\rm w}$ correlation with the inclusion of Swift 
GRBs for a
  wind circumburst medium case. The filled symbols (black circles and grey
  circles and stars) are the 25 GRBs considered for the fit of the
  correlation.
  Symbols are as in Fig. \ref {ghirla_homo}.}
\label{ghirla_wind}
\end{figure*}

\section{Results}

We first recomputed the $E_{\rm peak}$--$E_{\gamma}$ correlation in the case
of a homogeneous and wind--like medium and then the phenomenological \lz\ 
correlation following the method described in N06.

\begin{figure*}
\psfig{figure=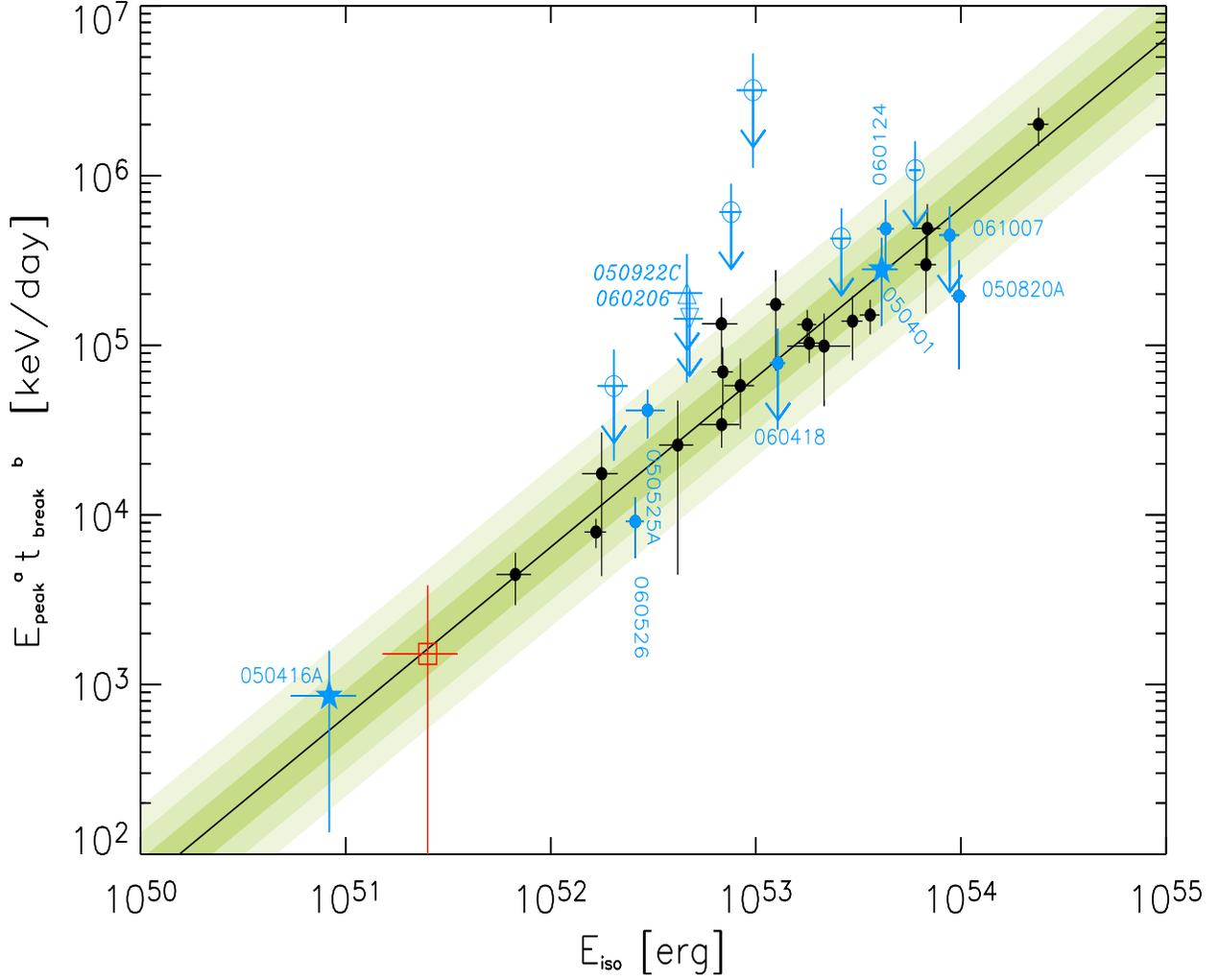,width=18cm,height=15cm}
\vskip -0.5 true cm
\caption{The phenomenological $E_{\rm peak}$-$E_{\rm iso}-t^{\prime}_{\rm jet}$ 
correlation with the inclusion of Swift GRBs. 
Symbols are as in Fig.\ref{ghirla_homo}.}
\label{lz}
\end{figure*}

To the pre--\sw\ bursts included in the N06 sample (17 GRBs -- excluding
GRB~050525A which is a \sw\ burst -- with measured jet break time) we add 16
new \sw\ GRBs.
The results using our total sample of 33 GRBs are presented in Fig.
\ref{ghirla_homo}, Fig. \ref{ghirla_wind}, and Fig. \ref{lz}: {\it all points
  and lower limits are consistent with these correlations and there are no
  outliers.}

Following the discussion of Sec. 2 and Sec. 3 we found 9 (out of 16) bursts
which have only a lower limit on their jet break (as discussed at the end of
Sec. 3 we include in the subsample of lower limits also GRB~050922C and
GRB~060206). Among these 9 lower limits there are 2 GRBs (i.e. 061007 and
060418) which lie on the right side of the \ghi\ correlation (see Figg.
\ref{ghirla_homo}, \ref{ghirla_wind}, \ref{lz}).  These 2 bursts were used for
the statistical analysis of the correlations.

In 7 cases (out of 16) the jet break was measured in the optical light curve
(in 5 cases $t_{\rm jet}$ was already published in the literature and in 2
bursts it is found for the first time in this work - Sec. 3).  However, we did
not consider in the statistical analysis GRB~060614. This bursts, in fact,
although fully consistent with both these correlations\footnote{ Note that the
  published spectral parameters of GRB~060614 make its peak energy poorly
  constrained.  This is the reason of the shown large uncertainties.  } (as
shown by the open square symbol in Figg. \ref{ghirla_homo}, \ref{ghirla_wind},
\ref{lz}), was claimed to represent a new class of bursts (see Gehrels et al.
2006; Della Valle et al. 2006b).  Therefore, the sample of \sw\ GRBs added to
the correlations and effectively used in their statistical analysis is
composed by 6 bursts with firm jet break measurements (GRB~050401, 050416A,
050525A, 050820A, 060124, 060526) and 2 bursts with lower limits on $t_{\rm
  jet}$ (GRB~060418 and 061007).  These, added to the pre--\sw\ sample of 17
bursts, bring to 25 the total number of GRBs used for the statistical
analysis.

We fitted the correlations by weighting for the errors on the involved
quantities (with the {\it fitexy} routine of Press et al. 1999).  
We also evaluated the scatter of the data points computed perpendicularly 
to the best fit line.

\begin{figure*}
\centerline{
\psfig{figure=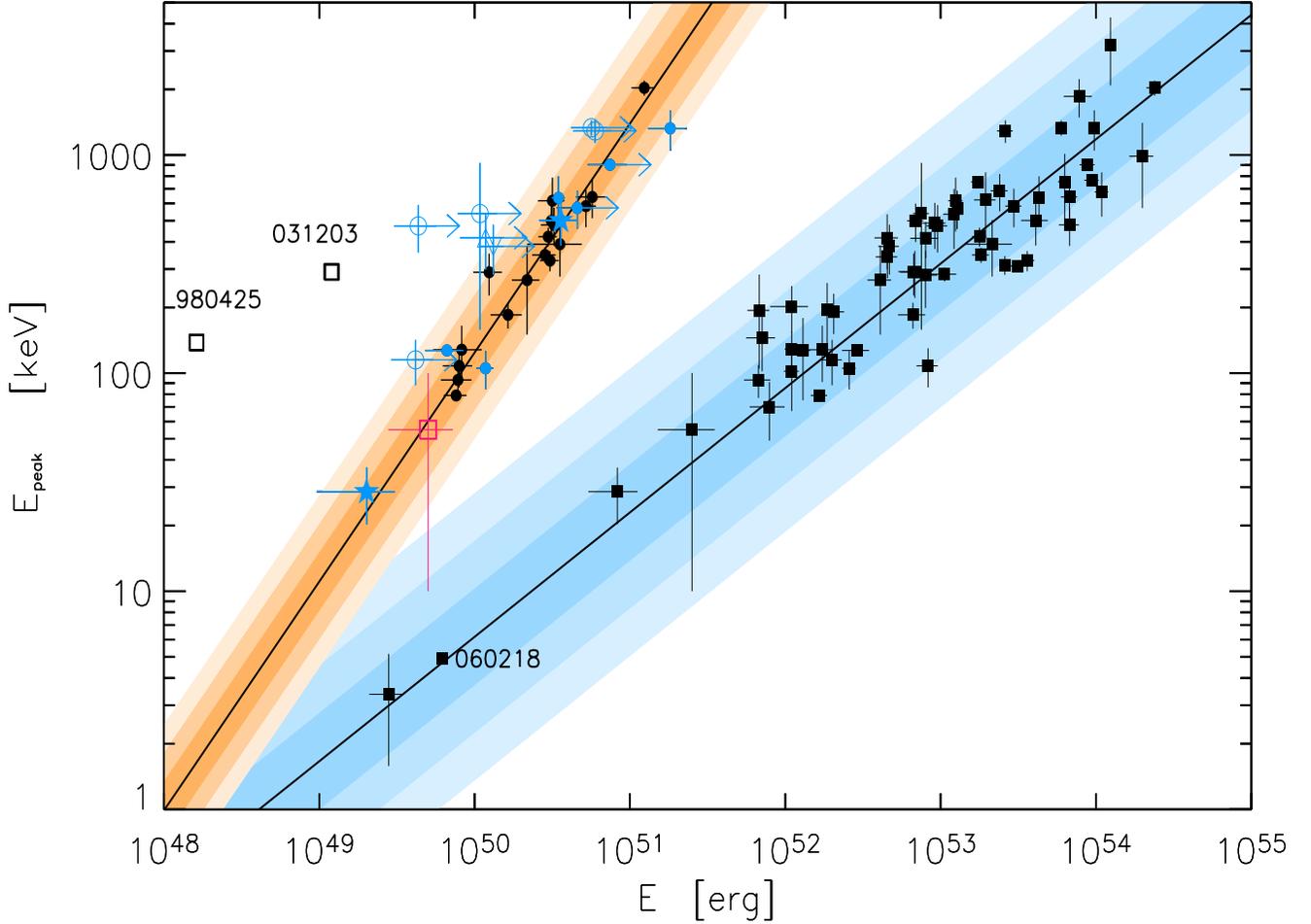,width=18.5cm,height=14cm}}
\vskip -0.5 true cm
\caption{The correlation between the rest frame peak energy and the bolometric isotropic
  energy (62 GRBs -- square symbols) with the most updated (up to Jan. 2007)
  burst sample is compared with the \ghi\ correlation (in the WM case). The
  two outliers GRB~980425 and GRB~031203 are shown. The solid lines represent
  the best fit to the data and the shaded regions the 1, 2, 3 $\sigma$ scatter
  of the data points around these correlations.}
\label{amaghi}
\end{figure*}

\subsection{The \ghi\ correlation in the HM case}

We recomputed the $E_{\rm peak}$--$E_{\gamma}$ correlation in the homogeneous
medium case. The jet opening angle is:
\begin{equation} 
\theta_{\rm jet}=0.161 \left({ t_{\rm jet,d} \over 1+z}\right)^{3/8} 
\left({n \, \eta_{\gamma}\over E_{\rm iso,52}}\right)^{1/8} 
\label{theta} 
\end{equation} 
where $z$ is the redshift, $\eta_\gamma$ is the radiative efficiency and
$t_{\rm jet,d}$ is the break time measured in days of the afterglow
light curve\footnote{
  Here we adopt the notation $Q=10^xQ_x$, and use cgs units
  unless otherwise noted.}.  
The efficiency $\eta_\gamma$ relates the
isotropic kinetic energy of the fireball after the prompt phase, $E_{\rm k,
  iso}$, to the prompt emitted energy $E_{\rm iso}$, through $E_{\rm k, iso}=
E_{\rm iso}/\eta_\gamma$. This implicitly assumes that $\eta_\gamma \ll 1$
otherwise the remaining kinetic energy after the prompt emission is instead
$E_{\rm k, iso}=E_{\rm iso} (1-\eta_\gamma)/\eta_\gamma$.  This efficiency, in
principle, could be different from burst to burst, but in the absence of any
hints of how its value changes as a function of other properties of the bursts
and favored by its low power in Eq. \ref{theta}, one assumes a constant value
for all bursts, i.e. $\eta_\gamma =0.2$ (after its first use by Frail et al.
2001, following the estimate of this parameter in GRB 970508).

The values of $\theta_{\rm jet}$ and $E_{\gamma}$ in the HM case are reported,
for all GRBs, in Tab. 1. With these values the \ghi\ correlation (shown in
Fig. \ref{ghirla_homo}) is
\begin{equation}
\left({E_{\rm peak} \over 100\, {\rm keV}}\right) \, =\, 
(3.02\pm0.14)\, 
\left({E_\gamma\over 4.4 \times 10^{50}\, {\rm erg}}\right)^{0.70\pm0.04}
\label{ghi_homo}
\end{equation}
with a reduced $\chi^{2}_{\rm r}=1.08$ for 23 degrees of freedom.  The errors
on its slope and normalization are calculated in the ``barycenter" of $E_{\rm
  peak}$ and $E_{\gamma}$, where the slope and normalization errors are
uncorrelated (Press et al. 1999). The scatter of the data points around this
correlation is distributed as a Gaussian with $\sigma=0.09$.

\subsection{The \ghi\ correlation in the WM case}

If the external medium is distributed with an $r^{-2}$ density profile
the semiaperture angle of the jet is related to the achromatic jet
break through (Chevalier \& Li 2000):
\begin{equation}
\theta_{\rm jet, w}\, = \, 0.2016 \, 
\left( {t_{\rm jet,d} \over 1+z}\right)^{1/4} 
\left( { \eta_\gamma\ A_* \over E_{\rm iso,52}}\right)^{1/4}
\label{thetaw} 
\end{equation}
where we assume $n(r)=Ar^{-2}$ and $A_*$ is the value of 
$A$ ($A=\dot M_{\rm w} /(4\pi v_{\rm w})=5\times 10^{11}A_*$ g cm$^{-1}$ ) 
when setting the mass
loss rate due to the wind $\dot M_{\rm w} =10^{-5} M_\odot$ yr$^{-1}$ and the
wind velocity $v_{\rm w}=10^3$ km s$^{-1}$, according to the typical
Wolf--Rayet wind
physical conditions. 
Given the few still uncertain estimates of the $A_*$
parameter, we assume a constant value (i.e. $A_*=1$) for all bursts
neglecting the unknown uncertainty on this parameter.

The fit with a power law model gives
\begin{equation}
{E_{\rm p} \over 100\, {\rm keV}}\, =\, 
(3.34\pm0.18)\, \left( {E_{\gamma, w} \over 2.6\times 10^{50}\, 
{\rm erg}}\right)^{1.05\pm 0.06}
\label{gglw}
\end{equation}
with a reduced $\chi^{2}_{\rm r}=0.89$ for 23 degrees of freedom (see Fig.
\ref{ghirla_wind}).  Note that the slope of this relation is entirely
consistent with unity.  The scatter of the points around the best fit
correlation is fitted by a gaussian with $\sigma=0.08$.

\subsection{The \lz\ correlation}

We have extended the 2D fit which weights for the errors on two quantities to
the 3D case for the \lz\ correlation. This fit weights the multidimensional
errors on the three independent variables $E_{\rm iso}$, $E_{\rm peak}$ and
$t_{\rm jet}$, where the peak energy and the jet break time are computed in
the source rest frame. We find
\begin{equation}
E_{\rm iso, 53} = (1.33\pm 0.10)
\left(E_{\rm peak} \over 299~ \rm keV \right)^{1.88\pm0.15}   
\left(t_{\rm jet} \over 0.53\rm d\right)^{-0.92\pm0.13}
\label{corr_fin}
\end{equation}
with a reduced $\chi^{2}_{\rm r}=1.33$ for 22 degrees of freedom. 
The scatter of the data points around the correlation is $\sigma=0.1$.

\subsection{The Amati versus Ghirlanda correlation}

Since we have discussed the most updated \ghi\ correlation we also compare it
with the most updated (up to Jan. 2007) $E_{\rm peak}-E_{\rm iso}$
correlation. Besides the GRBs in Tab.1 we have added those GRBs with $z$ and
$E_{\rm peak}$ known (excluding upper/lower limits) taken from the lists of
Amati (2006) and Nava et al.  (2007). We found a total of 64 GRBs. They are
shown in Fig.\ref{amaghi} (squares) with the \ghi\ correlation in the wind
case.  The fit of the $E_{\rm peak}-E_{\rm iso}$ correlation (using the {\it
  fitexy} routine and excluding the two outliers GRB~980425 and GRB~031203)
is:
\begin{equation}
{E_{\rm p} \over 100\, {\rm keV}}\, =\, 
(2.88\pm0.10)\, \left( {E_{\rm iso} \over 6.9\times 10^{52}\, 
{\rm erg}}\right)^{0.57\pm 0.01}
\label{ggal}
\end{equation}
with a reduced $\chi^2=7.2$ for 60 dof. The scatter of the data points around
the best fit is distributed as a Gaussian with $\sigma=0.2$.

From the comparison of the two correlations and their scatter it is noticeable
how the dispersion of the data points is reduced from the Amati to the
Ghirlanda relation, especially considering bursts with similar $E_{\rm peak}$.
Note that the different slopes of the Amati and Ghirlanda correlation imply
that they intersect at some small value of $E_{\gamma}=E_{\rm iso}$
corresponding to truly isotropic bursts. The intersection value, however,
depends on the circumbursts density profile: it is below $10^{48}$ erg in the
WM and smaller for the HM case.

\section{Discussion and conclusions}

In the era of multiple breaks of the X--ray and optical afterglow light curves
the identification of the jet break time is complex.  Even more so when flares
(more often in the X--rays, but sometimes also in the optical) occur, and the
lightcurve is not densely sampled.  In the following, for clarity, we list
some points which we believe are particularly important for the correct
identification of the jet break time.

\begin{itemize}
\item
The jet break should be observed in the optical.
\item
The light curve in the optical should extend
up to a time  longer than the jet break time
predicted by the \ghi\ correlation.
\item
The host galaxy and a possible supernova flux should be subtracted off.
\item
The break should be achromatic in the optical,
but we should relax the requirement of a simultaneous
break in the X--ray light curve, since the 
X--ray flux may be due to another component.
\item 
If a simultaneous X--ray and optical break is present,
one should check that this is not the time
break ending the plateau phase ($T_a$ in Willingale et al. 2007).
If this is the case, and the optical and X--ray flux continue
to track one another, it is possible that the component
dominating the X--ray flux is also contributing in the optical,
hiding any jet--break.
\item A steep [i.e. $F(t)\propto t^{-\alpha}$, with $\alpha>1.5$]
optical decay in a limited time interval is not necessarily an indication
of a jet break occurred earlier.
A steep decay can in fact be the tail of a flare. 
A densely sampled light curve is required to disentangle this case from
a post--break light curve.
\item The bolometric luminosity should be calculated
between 1 keV and 10 MeV in the rest frame of the GRB,
using the spectral parameters.
It is also useful, when a cut-off power law is used instead of the 
Band function, to give limits by using a Band model with a fixed 
$\beta$--value.
\item
Sometimes the Swift/BAT and the Konus--Wind data give different
spectral fits.
Given the limited energy range of BAT, one should take the
result of spectral fitting the BAT data with care.
\item An outlier for the \ghi\ correlation which is also an outlier for the
  $E_{\rm peak}$-$E_{\rm iso}$ correlation, while being an outlier, is
  ``recognizable" and should be put in the same category of GRB 980425 and GRB
  031203.
\item When deriving jet--angles from the jet break times, the same values of
  the efficiency $\eta$ should be used.  One should also allow for a range in
  possible densities of the interstellar medium (taken in the range 1--10
  cm$^{-3}$ in GGL04), when this is not derived by other means.  Of course
  this does not apply to the \lz\ correlation.
\end{itemize}

Following these rules we have selected, among all \sw\ long bursts with firm
redshift measurements (46 long GRBs up to Dec. 2006), those with measured peak
energy and some information about a jet break in the optical light curve.
%
%
We have found 16 \sw\ bursts that can be added to the pre--\sw\ sample of 17
GRBs (N06).  In the present sample of 33 events there is no outlier with
respect to the \ghi\ and \lz\ correlations (besides GRB 980425 and GRB 031203
-- but see Ghisellini et al. 2006 for the possibility that even these two GRBs
are not outliers).  
Indeed, these correlations are strengthened by the new
\sw\ bursts.  In particular we added 6 firm bursts (i.e. with defined 
$E_{\rm peak}$, $z$ and $t_{\rm jet}$) and 2 lower limits (on $t_{\rm jet}$) which
enlarge to 25 the original sample of pre--\sw\ busts used to fit these
correlations.  With this sample we re--analyzed the \ghi\ correlation in the
two possible scenarios of a homogeneous and wind medium (HM and WM). The
updated correlations are completely consistent with those found in the
pre--\sw\ era and their fits are statistically improved (i.e. $\chi^2=1.08$
and $\chi^2=0.89$ in the HM and WM case, respectively).  
We also re--analyzed
the phenomenological \lz\ correlation which turns out to be consistent with
what found before \sw\ (Liang \& Zhang 2005; N06).  The dispersion around
these correlations (computed with 25 GRBs) is described by a Gaussian with
$\sigma=$0.08--0.1 in all the three cases. Such dispersions are consistent
with being due to measurement errors only.

The confirmation of the \ghi\ and \lz\ correlations by the \sw\ GRBs also
supports the possibility, with an increasingly larger sample of
GRBs, to use them as standard candles to constrain the cosmological 
parameters (e.g. Ghirlanda, Ghisellini \& Firmani 2006).

The complexity of the afterglow light curves, disclosed by the \sw\ 
observations, certainly makes less direct and unambiguous the determination of
the jet break time, with respect to the pre--\sw\ era.  There are GRBs with
multiple breaks and flares, even in the optical band, and this makes mandatory
to have well sampled light curves beyond the jet break timescale predicted by
the \ghi\ and the \lz\ correlations.

\begin{acknowledgements} 
  We thank the referee, J. Norris, for his constructive and encouraging comments,
  and F. Tavecchio for useful discussion. This work made use of
  processed \sw\ data obtained from the public archive of N. Butler
  (http://astro.berkeley.edu/$\sim$nat/swift/). We thank a PRIN--INAF 2005 grant for
  funding.
\end{acknowledgements}

\end{document}